\documentstyle[12pt]{article}
\newcommand{\disc}{\displaystyle}

\newcommand{\lhisb}{{\cl}_{{\disc\chi}\mbox{sb}}}

\newcommand{\expon}[1]{{e}^{\disc{#1}}}

\newcommand{\vecr}{\vec{r}}

\newcommand{\fpi}{F_{\pi}}

\newcommand{\cl}{\cal {L}}

\newcommand{\apm}{{\hat{A}}^{+}_{1}{\hat{A}_2}}
\newcommand{\aone}{\disc{\hat{A}_1}}
\newcommand{\atwo}{\disc{\hat{A}_2}}
\newcommand{\uzero}{\disc{U_0}}
\newcommand{\uone}{{\disc}U_{1}}
\newcommand{\utwo}{\disc{U_2}}
\newcommand{\be}{\begin{equation}}
\newcommand{\ee}{\end{equation}}
\newcommand{\ba}{\begin{array}{l}}
\newcommand{\banonum}{\begin{eqnarray*}}
\newcommand{\ea}{\end{array}}
\newcommand{\eanonum}{\end{eqnarray*}}
\newcommand{\banum}{\begin{eqnarray}}
\newcommand{\eanum}{\end{eqnarray}}
\newcommand{\bb}{\begin{thebibliography}}
\newcommand{\eb}{\end{thebibliography}}
\newcommand{\ci}[1]{\cite{#1}}

\newcommand{\bi}[1]{\bibitem{#1}}

\newcommand{\Tr}{\mbox{Tr\,}}
\newcommand{\dd}{\partial}
\newcommand{\edc}{\end{document}}
\newcommand{\cg}{{C}_{\mbox{g}}}

\newcommand{\ga}{\disc{g_{A}}}

\newcommand{\mpi}{{\disc{m}_{\pi}}}

\newcommand{\rhoone}{{\disc\rho_{1}}}
\newcommand{\rhotwo}{{\disc\rho_{2}}}

\newcommand{\ltwo}{{\cl}_{\mbox{2}}}
\newcommand{\lw}{{\cl}_{\mbox{w}}}
\newcommand{\lfora}{{\cl}_{\mbox{4a}}}
\newcommand{\lfors}{{\cl}_{\mbox{4s}}}

\newcommand{\deltax}{\disc\delta_{x}}
\newcommand{\deltaz}{\disc\delta_{z}}
\newcommand {\bzero}{\disc{B_{0}}(\vecr)}
\newcommand{\riso}[1]{ {\langle {#1}^2 \rangle} }
\newcommand{\delxsq}{ \delta_{x}^{2} }
\newcommand{\delzsq}{ \delta_{z}^{2} }
\newcommand{\qzero}{ Q_{I=0} }
\newcommand{\qone} { Q_{I=1} }
\title{ The deformation  of the interacting
nucleon in the Skyrme model}
\author{A. Rahimov  \\
Theoretical   Physics Department, Tashkent State University \\
700095 Tashkent, Uzbekistan \\
T. Okazaki \\
  Physics Laboratory, Sapporo Campus,
        Hokkaido University of Education, \\
 Sapporo, Japan \\
M.M. Musakhanov\thanks{ E-mail: yousuf@phys.ualberta.ca; 
permanent address: Theor.Phys.Dept., Tashkent State
              University }, 
F. C. Khanna \thanks{ E-mail: khanna@phys.ualberta.ca}\\
Department  of Physics, University of  Alberta\\
Edmonton, Canada  T6G 2J1 \\
and\\
TRIUMF, 4004 Wesbrook Mall, \\
Vancouver, British Columbia, \\
Canada, V6T 2A3,\\}
\begin{document}\large
\begin{titlepage}
\maketitle
\begin{abstract}
Changes in the nucleon shape are  investigated by letting
the nucleon  deform under the strong interactions
with another nucleon. The parameters of
the axial deformations are obtained by minimizing the static energy of
the two nucleon system at each internucleon distance
$ R $. It is shown
that the intrinsic quadrupole moment of the interacting proton,  $Q_{p}$,
is about $0.02 fm^2$ at distances near $R \sim 1.25 $ fm.

\vskip 5mm
[ \bf\large{PACS numbers: 13.75 Cs , 11.10 Lm}; \\
 nucleon nucleon interaction,  nucleon shape, quadrupole moment ] 
 \end{abstract} 
\end{titlepage}

\newpage

 The possible changes in the radius of
 a nucleon in interaction
with another nucleon was investigated in the skyrmion model
by Kalbermann et. al.  \ci{kalberman1}.
It was found that there is a  swelling of the nucleon at intermediate
distances between  the nucleons which were assumed to preserve
their spherical shape. On the other hand
C. Hajduk and B. Schwesinger \ci{hs2}
considering the skyrmion
to be soft for  the deformation showed that there are several
deformed states of rotating skyrmions.
In particular,  besides the ground state of a spherical shape
there exists a degenerate doublet of exotic states with
the same quantum numbers as the nucleon ($ s=t=1/2 $) of oblate
and prolate shapes. One may therefore wonder whether the nucleon
can change its shape under the action of strong interactions.

Certainly, several studies have been made of the
deformation effects on the Skyrmions
\ci{oka3}, \ci{saito4}, \ci{kaulfus5}, 
but these were carried out to obtain a better
understanding of the possible sources of attraction in the central
nucleon - nucleon potential. As a result it was shown that the deformation
effect is very important and may reduce the central repulsion by
about $40\% .$

In a previous paper \ci{rm}, the problem of  the missing central attraction
in the $NN$ - interaction has been investigated within the model of Andrianov and Novozhilov \ci{anov6}, \ci{nikol7}
which starts with  the Skyrme lagrangian supplemented
 by a dilaton scalar field  $\sigma(r)$ so as to satisfy the QCD trace anomaly constraint. It was concluded that  this model   gives the desired attraction in the central part of the $NN$ interaction.
The resulting central scalar-isoscalar part of the potential
is in qualitative agreement with the phenomenological one.

In the present paper we shall concentrate
on the effects of the modification of the shape of a nucleon at a
 quantitative level
using skyrmions. We shall use the same model \ci{anov6}, \ci{nikol7}.
The lagrangian of this model including the chiral symmetry breaking term $\lhisb$ has the form:

\be
\ba
{\cl}(U,\sigma)=\ltwo(U,\sigma)+\lfora(U)+
\lw(\sigma) +\lhisb(U,\sigma) , \\
\ltwo(U,\sigma)=-{\disc\frac{\fpi^{2}}{16}}[{\Tr}L_{\mu}L^{\mu}-2{(\dd_{\mu}
{\sigma})}^{2}]{e}^{-2{\disc\sigma}}, \\
{\lfora}(U)={\disc\frac{1}{32{e}^{2}}}{\Tr}{[L_{\mu} , L^{\nu}]}^{2} , \\
\\
\lw(\sigma)=-{\disc\frac{\cg}{24}}[\expon{-4\sigma}-1
+{\disc\frac{4}{\varepsilon}} (1-\expon{-{\varepsilon}\sigma})]   \\
{\lhisb}(U,\sigma)={\disc\frac{{e}^{-3{\disc\sigma}}{\mpi}^{2}{\fpi}^{2}}{8}}\!\Tr(U-1)\\
\ea
\ee

where ${\fpi}$ is the pion decay constant,
$\mpi$ is mass of the pion,  $ e$  is the coupling constant
of the  Skyrme term $\lfora$ , $\cg$ is gluon
condensate parameter   $\sigma$ is scalar meson field ,
$U$ is  an $SU(2)$ matrix chiral field and
$L_{\mu}=U^{+}\dd_{\mu}U . $
It is a generalization of the well - known original Skyrme model
\ci{skyrme8}, \ci{anw9}
and takes into account the conformal anomaly
of the QCD.
The term $  \ltwo$ includes the kinetic term of the chiral  and
scalar fields. The effective potential for the scalar field
was calculated by Migdal and Shifman \ci{migdal10}.
The parameter $\varepsilon$
depends on the number of flavors $N_f$:
 $ \varepsilon=8N_{f}/(33-2N_f)$.
Note that in the limit of a heavy $\sigma$-- meson the potential
becomes equal to the symmetric quartic term
  $\:\:\:{\lfors}={\disc\frac{\gamma}{8e^{2}}}
{[{\Tr}L_{\mu}L^{\mu}]}^{2} ,\:\:\:$ 
which is necessary to reproduce the $\pi\pi$
scattering data \ci{donoghue11}.
The main difference between the  model of refs.     \ci{gomm12},
\ci{kalberman1}, \ci{yabush13}, \ci{wambach14}
and this one (eq.(1)) is in the origin of
the dilaton. In the former  \ci{gomm12}
the  dilaton is associated
with the glueball while in the latter \ci{anov6}
it is associated with  the
quarkonium. Nevertheless  both models produce
 in a natural and transparent way the intermediate - range
  attraction in the
 central part of baryon - baryon interaction
 even in the product approximation  \ci{yabush13}.
We assume that both skyrmions deform into an ellipsoidal shape
 as they come  close together. To describe this we write the chiral field
 $U$ and the dilaton field $\sigma$ as the nonspherical hedgehog
 form given by :
 \be
 \ba
  U_{0}(\vecr)=\exp{ (i\vec{\tau}\hat{q}\Theta(q))}\\
  \sigma(\vecr)=\sigma(q)
 \ea
 \ee
 where the spatial vector $\vec{q}$ has the components
 ${\deltax}x   $, ${\deltax} y$,${\deltaz} z,$ 
 $\hat{q}$ is the unit vector: $\hat{q}=\vec{q}/q$
 with the deformation parameters $\deltax$ and $\deltaz$.
 The profile functions  $\Theta$ and $\sigma$ are assumed to be
 the solutions of the Euler - Lagrange equations in the
 spherical case given by \ci{saito4}.
This ansatz,  eq. (2), leads to a  modification of
 the static mass,  $\disc{M}^{*}_{H}$,
  and the moment of
 inertia,  $\disc{\lambda}^{*}_{M}$, of the Skyrmion
 \be
 \ba
 \disc{M}^{*}_{H}=[AM_{2}+BM_{4a}+M_{{\chi}sb}+M_{W}]/\eta\\
 \disc{\lambda}^{*}_{M}=[\lambda_2+A\lambda_{4a}]/\eta
 \ea
\ee
where $\eta=\deltax^2\deltaz$, $A=(2\deltax^2+\deltaz^2)/3$,
$B=\deltax^2(\deltax^2+2\deltaz^2)/3$
and  $M_i$ and $\lambda_i$ denote the relevant contributions
from ${\cl}_{i} $ term in eq. $(1)$  for the spherical case
$\deltax=\deltaz=1$.
As we are mainly interested in the region where the medium
range attraction takes place - $R\sim 1.25 fm$
( typical separation between
nucleons in  nuclei)  we restrict ourselves to  the familiar
product ansatz:
\be
\ba
U =
\aone{\uzero}(\vec{X}-\vec{q}/2)\apm{\uzero}(\vec{X}+\vec{q}/2)
{\atwo^{+}}\equiv
\uone\utwo \\
{\rho}=\rho(\vec{X}-\vec{q}/2)\rho(\vec{X}+\vec{q}/2)
\equiv\rhoone\rhotwo \;\; , \;\rho\equiv\exp(-\sigma)\\
\ea
\ee
where $\aone$, $\atwo$ are the collective coordinates
of skyrmions to describe
their rotational  motion,
 $\vec{q} $ is the vector along $z$ axis: $q_x=0,$
$q_y=0,$ $q_z=q=R\deltaz $ and  $R$ is the distance between skyrmions.
The static skyrmion - skyrmion potential is defined by:
\be
V(\vec{R},\deltax,\deltaz)=
-{\int}d\vec{X}[{\cl}(\uone\utwo , \rhoone\rhotwo)-{\cl}(\uone , \rhoone)
-{\cl}(\utwo , \rhotwo)]
\ee
The application  of   the usual projection methods
 developed in \ci{jjp15}, \ci{otof16}, \ci{depace17}, \ci{nyman18},
 \ci{yabuando19}
to eq.s $(1)-(5)$ yields the following representation
 for the
  central scalar - isoscalar part of the nucleon - nucleon interaction :
 \be
 V_{c}(R,\deltax,\deltaz)=\disc\frac{1}{\eta}
 [V_{{\chi}sb}(q)+V_{W}(q)+
 \deltax^2(V_2(q)+\deltaz^2 V_{4a}(q))+(\deltaz^2-\deltax^2)(V_{2}^{def}(q)+\deltax^2 V_{4a}^{def}(q))]
 \ee
where the terms $ V_{2}^{def} $ and  $ V_{4a}^{def} $
 are the net contributions
from the deformation effect of  the terms
 $\ltwo$ and $\lfora$ in eq.(1) respectively. The deformation parameters
 $\deltax(r) $ and $\deltaz(r)$ were calculated by minimizing
 the total static energy of the two nucleon system at each separation
 $R$ using eq. $(2) $ and eq. $(6)$.

 The resulting values of the parameters are
 then used to study the changes in the
 shape of the nucleon. We will illustrate this
 procedure for the case of the  isoscalar mean square radius
  and the appropriate intrinsic quadrupole moment.
 The normalized isoscalar mean square radius
  along each axis may be defined by:
\be
{ \riso{r_i} }^{*}_{I=0}=\disc\frac{ {\int}d{\vecr}r_{i}^{2}\bzero }
{{\int}d{\vecr}\bzero }
\ee
where $(i=x,y,z) ,$  $\bzero$ is the baryon charge distribution
\be
\bzero=\frac{1}{24{\pi}^{2}}{\epsilon}^{ijk}{\Tr}[L_{i}
L_{j}L_{k}].\\
\ee
The inclusion of the deformation in a simple way :
 $ r_{i}{\rightarrow}q_{i}/\delta_{i} $ as in eq. $(2)$
 yields the following relation between the radius of a free
 spherical nucleon     ${\riso {r}}_{I=0} $ and a deformed one :
 ${ \riso{r_i} }^{*}_{I=0}=\disc\frac{ 1 } {\delta_{i}^{2} }
 {\riso {r}}_{I=0}
 $ .
 Therefore the appropriate
  quadrupole moment characterizing the shape of the
baryon matter distribution is compared  to that of an ellipsoid with axis
$1/\delta_{z}$ and $1/\delta_x :$
 $  Q_{I=0}=3 { \riso{r_z} }^{*}_{I=0}
-{\riso{r}}^{*}_{I=0}=2{\riso{r}}_{I=0}(1/\delzsq-1/\delxsq)$.
The explicit formulas  for $Q_{I=1}$ defined by
 $  Q_{I=1}=3 { \riso{r_z} }^{*}_{I=1}
-{\riso{r}}^{*}_{I=1}$
 are rather complicated and may by found elsewhere \ci{jafsubmit20}.

 In the numerical calculations we consider the following two
 cases: the lagrangian with the dilaton and the pure Skyrme model
 when $\sigma=0$ in eq. (1). In  both cases the parameters $\fpi ,$
 $e$ and $\mpi$ were fixed at the values:
  $\fpi=186 MeV,$  $ e=2\pi ,$  $\mpi=139MeV.$ For the gluon condensate
  we use $\cg=(283 MeV)^4$ as obtained from lattice QCD calculations
\ci{satz21}.
The mass of the scalar meson,  $m_{\sigma}$,  defined by
$ m_{\sigma}=\sqrt{2\cg}/\fpi$ is then $ 610 MeV$.

This set of parameters produce the following
  static properties of the nucleon: $M_N=1054 MeV$,
  $\ga=0.65$, $ { \riso{r} }^{1/2}_{I=0}=0.38 fm $  and
  $ { \riso{r} }^{1/2}_{I=1}=0.66 fm $    in the dilaton case.
  No attempt is made here to search for a realistic set of
parameters
  since our interest is mainly to establish a link between the
properties  of $NN$ interaction and the shape of the nucleon.

In Fig.1 
 the central scalar - isoscalar part of the nucleon - nucleon interaction
 has been presented for the two cases (both including deformation effects) with  the dilaton field (solid curve) and without one (dashed curve).
 For  comparison 
the realistic phenomenological  "Paris" potential \ci{paris22}
 is also  displayed ( the dotted  line in Fig.1).
 It is clear that  the lagrangian with the dilaton field
is able to    describe the nucleon - nucleon interaction in the
   intermediate region quite well.

In the case with the dilaton field, the deformation effects give a contribution to the the central scalar - isoscalar part of the nucleon - nucleon potential $ \sim -2MeV$ at  the minimum point $R\sim1.25fm$.
It means that  an  attraction in the central part of $NN$ interactions is provided mainly by the dilaton field.

In addition to the static (adiabatic) potential, there are also dynamical  $R$ dependent effects in the reduced mass  of 
$NN$ system.  This dependence gives rise to a velocity dependent attraction. However, it is gives a small contribution at  low energies   \ci{oka}.

    We now turn to changes in the shapes of the  interacting nucleons.
    The intrinsic quadrupole moments
     $\qzero$ and $\qone$ are shown in  Fig.2
    and Fig.3 respectively.In the pure Skyrme model
     when there is no attraction ($V_2=V_{2}^{def}=0$ in eq. (6))
     between skyrmions it  becomes oblate  (dashed lines in the Figures 2, 3)
     due to  the strong repulsion caused by the $V^4$ terms
     in eq. (6). The inclusion of the dilaton leads to the following
     qualitative picture: At  large separations skyrmion is
     obviously in a spherical shape, becomes prolate
      at the intermediate region and deforms to an oblate shape
at small        distances where the repulsion dominates.
      As the nucleons approach each other they change shapes from prolate
      into   oblate       at $R\sim1.2fm$.Comparing  Figures 2 and 3
      it  may be noticed  that the isoscalar intrinsic quadrupole moment $\qzero$
      is much smaller than the isovector one $\qone$ at
      intermediate  separations.

     The intrinsic quadrupole moment of the proton  defined by:
     $ Q_{p}=(\qzero+\qone)/2 $ reaches a  maximum value of
     $Q_{p}=0.016 fm^2$
     at $r\sim 1.5 fm.$  Hence, one may conclude
     that the shape of a nucleon in  nuclei is not
     spherical. We expect new data from high - energy electron
     scattering on nuclei to make this situation  clear.

  As a concluding remark we have to underline that the deformed states
 of oblate (prolate) shapes may  not necessarily belong to
 the $K=1$ band  \ci{hs2} since for a strongly
 deformed system  the quantization procedure
 used here needs some modifications.

This work was supported in part  by International Science Foundation
Grant MZJ000 and JSPS grants. The work of F.Khanna is
supported in part by the Natural Sciences and Engineering research Council
of Canada.

\bb{99}
\bi{kalberman1}G. Kalbermann, L. L. Frankfurt and J. M. Eisenberg, 
{\it Phys. Lett.} {\bf B329} (1994) 164.
\bi{hs2}C. Hajduk and B. Schwesinger, 
{\it Nucl. Phys. } {\bf A453} (1986) 620.
\bi{oka3}M. Oka, 
{\it Phys. Rev.} {\bf C36} (1987) 720.
\bi{saito4}S. Saito, 
{\it Progr. Theor. Phys. (Supplement)} {\bf 91} (1987) 181.
\bi{kaulfus5}U. B. Kaulfuss and U. -G. Meissner, 
{\it Phys. Rev.} {\bf D31} (1985) 3024.
\bi{rm}M.M. Musakhanov  and  A.M. Rahimov, 
{\it Mod. Phys. Lett.} {\bf A10} (1995) 2297.
\bi{anov6} V. Andrianov and V. Novozhilov, 
{\it  Phys. Lett.} {\bf B202} (1988) 580.
\bi{nikol7}V. A. Nikolaev, V. Novozhilov and O. G. Tkachev,
{\it Preprint of  JINR} (Rapid communications ) {\bf 2 [53] - 92} (1992).
\bi{skyrme8} T. H. R. Skyrme, 
{\it Nucl. Phys.} {\bf 31} (1962) 556.
\bi{anw9} G. S. Adkins, C. R. Nappi and E. Witten,
 {\it Nucl. Phys.} {\bf  B228} (1983) 552.
\bi{migdal10}A. Migdal and M. Shifman,  
{\it Phys. Lett. } {\bf B114 } (1982) 445.
\bi{donoghue11}J. Donoghue, E. Golowich and B. Holstein,
 {\it Phys. Rev. Lett.} {\bf 53} (1984) 747.
\bi{gomm12}H. Gomm, P. Jain, R. Johnson and J. Schechter,
 {\it Phys. Rev.} {\bf D33} (1986) 819; ibid   3476.
\bi{yabush13} H. Yabu, B. Schwesinger and G. Holzwarth, 
{\it Phys. Lett.} {\bf B224} (1989) 25.
\bi{oka}M. Oka, 
{\it Nucl. Phys.} {\bf A463} (1987) 247c.
\bi{wambach14}T. Waindzoch and J. Wambach, 
{\it Phys. Lett.} {\bf B295} (1992) 16.
\bi{jjp15}A. Jackson, A. D. Jackson and V. Pasquer, 
{\it Nucl. Phys.} {\bf A432} (1985) 567.
\bi{otof16}T. Otofuji et al., 
{\it Phys. Rev.} {\bf  C34} (1986) 1559.
\bi{depace17}A. De Pace, H. M{\"u}ther and A. Faessler,
 {\it Z. Phys.} {\bf A325} (1986) 229.
\bi{nyman18}E. M. Nyman and D. O. Riska,
 {\it Phys. Scripta} {\bf 34} (1986) 533.
\bi{yabuando19}H. Yabu and K. Ando, 
{\it Progr. Theor. Phys.} {\bf 74} (1985) 750.
\bi{jafsubmit20}M. M. Musakhanov and A. Rahimov, 
submitted  to { \it Yad. Phys.} (1995).
\bi{satz21}H. Satz, 
{\it Phys. Lett.} {\bf B113} (1982) 245.
\bi{paris22}M. Lacombe et al.,
{\it Phys. Rev.} {\bf C21} (1980) 861.
\eb

\newpage
\centerline {\bf Figure captions}
\begin{description}
\item [Fig.1.]
Central isospin - independent potential (eq. 6)
calculated for the cases with  the dilaton field
(solid curve) and without one (dashed curve) as
a function of the internucleon distance $R$ .
The dotted curve is the corresponding interaction component
in the "Paris" potential \ci{paris22}.
\item [Fig.2.]
The  $R$ dependence of the isoscalar quadrupole moment
$\qzero$ of the nucleon.The solid and dashed lines
are obtained in the case with dilaton  and pure Skyrme
model respectively.
\item[Fig.3.]
The same as in Fig.2 but for the isovector  quadrupole moment
$\qone$.
\end{description}
\newpage

\let\picnaturalsize=N
\def\picsize{8.0in}
\def\picfilename{fig1.eps}
\ifx\nopictures Y\else{\ifx\epsfloaded Y\else\input epsf \fi
\let\epsfloaded=Y
\centerline{\ifx\picnaturalsize N\epsfxsize \picsize\fi \epsfbox{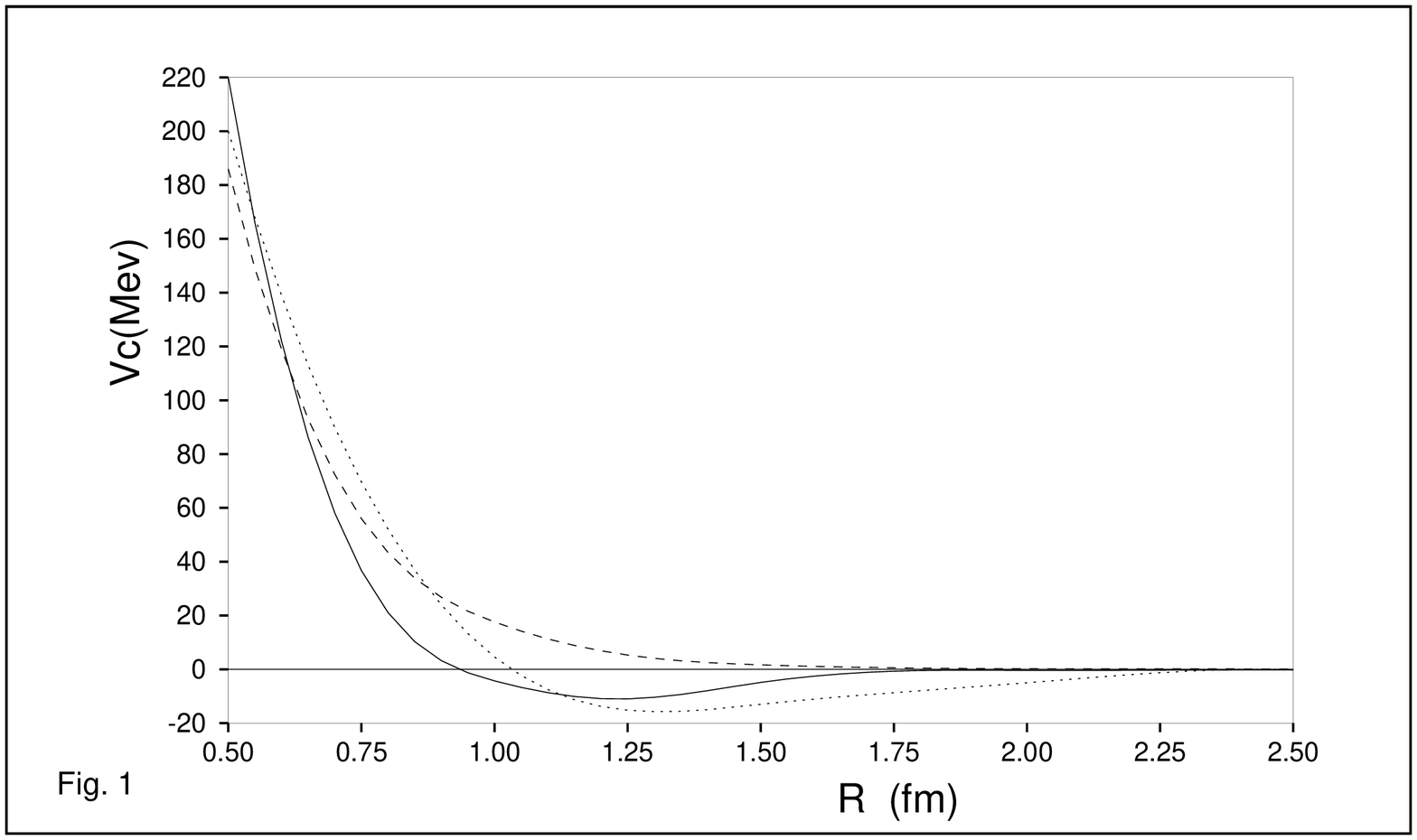}}}\fi
\newpage

\let\picnaturalsize=N
\def\picsize{8.0in}
\def\picfilename{fig2.eps}
\ifx\nopictures Y\else{\ifx\epsfloaded Y\else\input epsf \fi
\let\epsfloaded=Y
\centerline{\ifx\picnaturalsize N\epsfxsize \picsize\fi \epsfbox{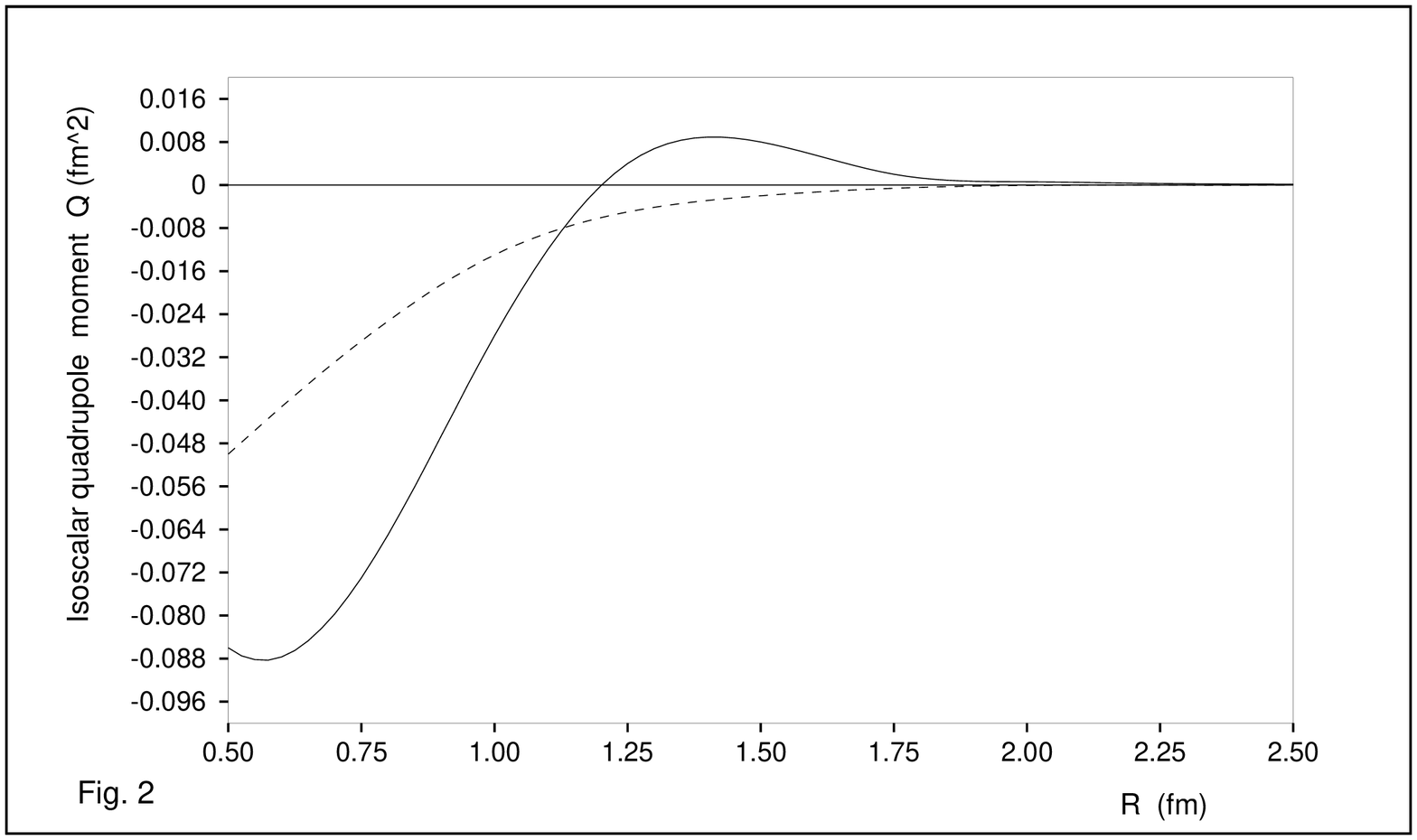}}}\fi
\newpage

\let\picnaturalsize=N
\def\picsize{8.0in}
\def\picfilename{fig3.eps}
\ifx\nopictures Y\else{\ifx\epsfloaded Y\else\input epsf \fi
\let\epsfloaded=Y
\centerline{\ifx\picnaturalsize N\epsfxsize \picsize\fi \epsfbox{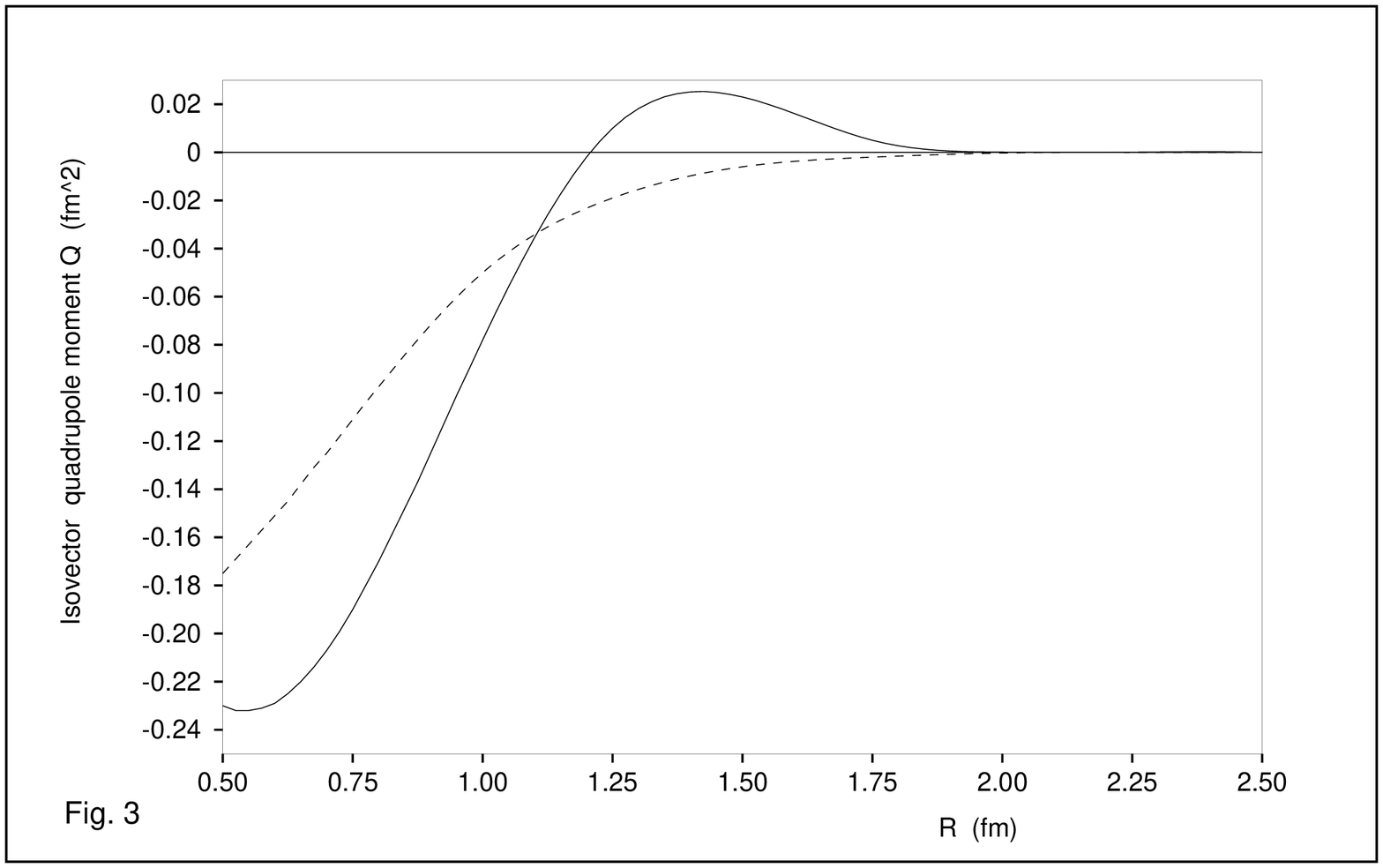}}}\fi

\edc